\documentclass[showpacs,preprintnumbers,amsmath,amssymb,twocolumn,superscriptaddress,prb]{revtex4}
\usepackage{times}
\usepackage{graphicx}
\usepackage{amsfonts}
\usepackage{amsmath, amsthm, amssymb}
\usepackage{dsfont}

\newcommand{\ve}[1]{\mathbf{#1}}
\newcommand{\mf}{m_\text{F}} 
\newcommand{\ms}{m_\text{S}} 
\newcommand{\vk}{\ve{k}} 

\newcommand{\e}[1]{\mathrm{e}^{#1}}

\newcommand{\ie}{\textit{i.e. }}
\newcommand{\eg}{\textit{e.g. }}
\newcommand{\etal}{\emph{et al.}}
\def\i{\mathrm{i}}

\begin{document}
\title[Signatures of retroreflection and induced triplet electron-hole correlations in ferromagnet/$s$-wave 
superconductor structures]{Signatures of retroreflection and induced triplet electron-hole correlations in ferromagnet/$s$-wave 
superconductor structures}
\author{J. Linder}
\affiliation{Department of Physics, Norwegian University of
Science and Technology, N-7491 Trondheim, Norway}
\author{A. Sudb{\o}}
\affiliation{Department of Physics, Norwegian University of
Science and Technology, N-7491 Trondheim, Norway}
\affiliation{Center for Advanced Study, The Norwegian Academy of 
Science and Letters, N-0271 Oslo, Norway}

\date{Received \today}
\begin{abstract}
We present a theoretical study of a ferromagnet/$s$-wave superconductor junction to investigate the signatures of induced triplet correlations in the system. We apply the extended BTK-formalism and allow for an arbitrary magnetization strength/direction of the ferromagnet, a spin-active barrier, 
Fermi-vector mismatch, and different effective masses in the two systems. It is found 
that the phase associated with the $xy$-components of the magnetization in the ferromagnet
couples with the superconducting phase and induces spin-triplet pairing correlations
in the superconductor, if the tunneling barrier acts as a spin-filter. This feature leads 
to an induced spin-triplet pairing correlation in the ferromagnet, along with a spin-triplet 
electron-hole coherence due to an interplay between the ferromagnetic and superconducting 
phase. As our main result, we investigate the experimental signatures of retrorelection, manifested in the 
tunneling conductance of a ferromagnet/$s$-wave superconductor junction with a spin-active interface. 
\end{abstract}

\maketitle

\section{Introduction}
\noindent The proximity effect \cite{buzdinRMP} in a normal/superconductor (N/S) junction 
refers to the induced superconducting correlations between electrons and holes in the normal 
part of the system. Even far away from the junction (typically distances much larger than 
the superconducting coherence length $\xi$) where the pairing potential is identically equal 
to zero, these correlations may persist. Consequently, the proximity effect is responsible 
for a plethora of interesting physical phenomena, including the 
Josephson effect in S/N/S junctions \cite{kulik}, the spin-valve effect in ferromagnet/superconductor 
(F/S) layers \cite{buzdin1999}, and the realization of so-called $\pi$-junctions, which in particular 
have received much attention both theoretically \cite{pitheoretical} and experimentally \cite{piexperimental} 
during the past decades. The understanding of Andreev-reflection processes \cite{andreev1964} is crucial 
when dealing with the proximity effect in N/S systems. Roughly speaking, this phenomenon may be thought 
of as a a coherently propagating electron with energy less than the superconducting gap $\Delta$ 
incident from the N side of the barrier being reflected as a coherently propagating hole, while in 
the process generating a propagating Cooper pair in the S. Such processes are highly relevant in 
the context of transport properties of N/S heterostructures in the low-energy regime, and have 
proved to be an effective tool in probing the pairing symmetry of unconventional SCs (see
Ref.~\onlinecite{deutscherRMP} and references therein). \\
\indent In recent years, the fabrication of ferromagnet/superconductor heterostructures has been 
subject to substantial advances due to the development of techniques in material growth and high 
quality interfaces \cite{weider, wei}. With an increasing number of recently discovered unconventional 
superconductors with exotic pairing symmetries \cite{saxena, bauer, akasawa}, there exists an 
urgent need to refine the traditional methods, such as tunneling spectroscopy, in order to 
correctly identify the experimental signatures which reveal the nature of the pairing 
potential for such superconductors. For one thing, this amounts to taking into account effects 
which are known to be present in tunneling junction experiments and that may significantly 
influence the conductance spectra, such as local spin-flip processes and the non-ideality 
of the interface \cite{kreuzer}. Also, with the aim of producing theoretical tools that may 
serve as a guide for identifying the superconducting pairing symmetry, possible spin-filter 
effects of interface in ferromagnet/superconductor heterostructures warrant attention 
\cite{garcia}.\\
\indent Studies of quantum transport in F/S junctions have a 
long tradition for both conventional and unconventional pairing symmetries in the superconductor 
\cite{jong, zhu, zutic99, kashiwaya1999}. Currently, such systems have become the subject of 
much investigation, not only due to their interesting properties from a fundamental physics 
point of view, but also because such heterostructures may hold great potential 
for applications in nanotechnological devices. An important characteristic of most F/S junction 
is that, unlike N/S junctions, \textit{retro-reflection} is absent for the hole in the F part 
of the system. This means that the reflected hole, which carries opposite spin of the original 
electron, does not retrace the trajectory of the incoming electron. The absence of 
retro-reflection is due to the presence of an exchange interaction. Previous studies of such 
systems have primarily focused on a magnetization lying in the plane of the F/S junction, where 
in most cases the barrier contains a pure non-magnetic scattering potential 
\cite{jong, zhu, zutic99}. Kashiwaya \etal\; \cite{kashiwaya1999} included the effect 
of a magnetic scattering potential in this type of junction, \ie spin-active barriers, 
and very recently, it was suggested by Kastening \etal\; \cite{kastening} that the 
presence of both intrinsic and spin-active scattering potentials in the barrier of a S/S 
junction may lead to qualitatively new effects for the Josephson current. So far, 
the influence of the F phase associated with the planar magnetization perpendicular 
to the interface has been largely unexplored, although Ref.~\onlinecite{kastening} considers 
the 1D case of this situation. \\
\indent It is therefore the purpose of this paper to investigate two interesting 
features that arise in a F/S junction in the presence of planar magnetization components: 
\textit{i)} the interplay between the planar magnetization and the presence of a 
spin-active barrier may restore retro-reflection for a given parameter range, and 
\textit{ii)} the resulting induced electron-hole pair correlations exhibit a coupling 
between $\phi$ and the S phase $\gamma$. Since our findings suggest that the traditional 
picture of absent retro-reflection does not hold for planar magnetization with respect 
to the junction in the presence of a spin-active barrier, we argue that these results 
are of major importance in the study of F/S junctions. The presence of retro-reflection 
in a F/S junction thus influences the spin-charge dynamics in a significant way, giving 
rise to new possibilites of quantum transport involving charge- and spinflow in such a 
heterostructure. Elucidating the consequences of this is of fundamental importance. It 
is also of considerable importance in device fabrication, since our results imply that 
the spin-active properties of a tunneling barrier play a crucial role.    
\begin{figure}[h!]
\centering
\resizebox{0.47\textwidth}{!}{
\includegraphics{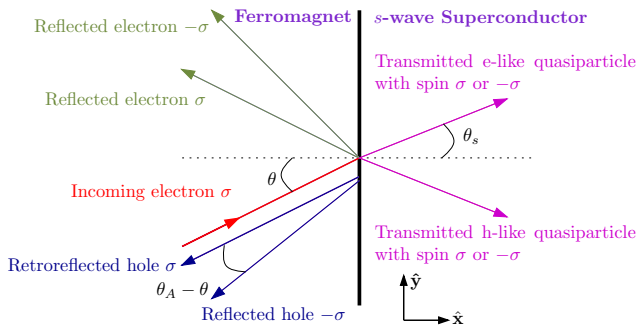}}
\caption{(Color online). Schematic overview of the relevant scattering processes that 
take place at the F/S interface. We take into account the possibility of retroreflected 
holes with equal spin as the incoming electron. This is due to the presence of spin-flip 
processes manifested in the form of planar magnetization and a spin-active barrier.}
\label{fig:model}
\end{figure}
\\
This paper is organized as follows. In Section II, we define the model we 
study and set up definitions of the scattering amplitudes to be considered. 
In Section III we investigate what conditions are necessary for retroreflection 
to occur. In Section IV, we give our results for the conductance. In Section IV A, we consider the 
influence of Fermi-vector mismatch on the conductance spectrum $G(E)$, in 
Section IV B we consider the effect of exchange energy on $G(E)$, in Section 
IV C we consider the effect of differing effective masses across the tunneling 
junction on $G(E)$, and in Section IV D we consider the effect on $G(E)$ of 
varying the relative strength of magnetic and non-magnetic scattering potential 
in the contact region between F and S. In Section V we provide a discussion
of results, including a comparison of our results to earlier ones on similar 
problems. We highlight what our new findings are compared to earlier results. 
Finally, Section VI summarizes our results. 

\section{Model and formulation}
\noindent We define our model as follows. Consider a 2D F/S junction as illustrated 
in Fig. \ref{fig:model}. As is seen from the figure, $\theta$ is the angle of incidence 
for electrons with spin $\sigma$ that feel a barrier strength 
$V_\sigma(x) = (V_0 - \sigma V_s)\delta(x)$, where $V_0$ and $V_s$ is the non-magnetic 
and magnetic scattering potential, respectively, \ie the barrier is spin-active 
\cite{kashiwaya1999}. Physically, this means that the barrier acts as a spin-filter.
Furthermore, $\theta_A$ is the angle of reflection for particles with spin 
$-\sigma$. The Bogoliubov de Gennes (BdG)-equations that describe the quasiparticle 
states $\Psi(x,y)$ with energy eigenvalues $E$ in the two subsystems are given by 
\begin{align}
\begin{pmatrix}
\hat{H}_0(x,y) & \hat{\Delta}(x)\\
-\hat{\Delta}^\dag(x) & -\hat{H}_0^\text{T}(x,y)\\
\end{pmatrix}
\Psi(x,y) = E\Psi(x,y),
\end{align}
where we have defined the single-particle Hamiltonian
\begin{align}
\hat{H}_0(x,y) &= -\boldsymbol{\nabla}_{xy}^2/[2m_\text{F}\Theta(-x) + 2m_\text{S}\Theta(x)] \notag\\
&- \boldsymbol{\hat{\sigma}}\cdot\boldsymbol{M}\Theta(-x) + \text{diag}(V_\uparrow(x),V_\downarrow(x)),
\end{align} while $\hat{\Delta}(x) = \i\hat{\sigma_y}\Delta(x)$. We allow for different effective 
masses in the two systems, given by $m_\text{F}$ and $m_\text{S}$. The magnetic exchange energy 
splitting is denoted 
\begin{equation}
M = (M_{xy}^2 + M_z^2)^{1/2},
\end{equation}
where $M_{xy}^2 = M_x^2 + M_y^2$ is the planar contribution of the magnetic exchange 
energy, while $2M_z$ is the energy-splitting between spin-$\uparrow$ and spin-$\downarrow$ 
bands. The quasiparticle wave-vectors are then given by 
\begin{align}
k^\sigma &= \sqrt{2m_\text{F}(E_\text{F} + \sigma M)},\notag\\
q &= \sqrt{2m_\text{S}E_\text{S}}
\end{align}
in the F part and S part of the system, respectively, where $\mu_i$ is the chemical 
potential. We have made use of the standard approximation $\mu_i \gg \Delta$. Moreover, 
we take the S order parameter to be constant up to the junction such that 
$\Delta(\gamma,x) = \Delta\e{\i\gamma}\Theta(x)$. Solving the BdG-equations, the 
wave-functions $\psi$ on the F side and $\Psi$ on the S side become
\begin{widetext}
\begin{align}\label{eq:wave}
\psi(x,y) &= \e{\i k_yy}\Bigg[ 
\begin{pmatrix}
s_\uparrow a\\
s_\uparrow b\e{-\i\phi}\\
0\\
0\\
\end{pmatrix}
\e{\i k^\uparrow \cos\theta x} + 
\begin{pmatrix}
-s_\downarrow b\e{\i\phi}\\
s_\downarrow a\\
0\\
0\\
\end{pmatrix}
\e{\i k^\downarrow \cos\theta x} +
r_e^\uparrow\begin{pmatrix}
a\\
b\e{-\i\phi}\\
0\\
0\\
\end{pmatrix}
\e{-\i k^\uparrow S x}\notag\\
&\hspace{0.3in}+ r_e^\downarrow\begin{pmatrix}
-b\e{\i\phi}\\
a\\
0\\
0\\
\end{pmatrix}
\e{-\i k^\downarrow \tilde{S}x}+
r_h^\uparrow\begin{pmatrix}
0\\
0\\
a\\
b\e{\i\phi}\\
\end{pmatrix}
\e{\i k^\uparrow S x}
+ r_h^\downarrow\begin{pmatrix}
0\\
0\\
-b\e{\i\phi}\\
a\\
\end{pmatrix}
\e{\i k^\downarrow \tilde{S} x}\Bigg]\notag\\
\Psi(x,y) &= \e{\i k_yy} \Bigg[
t_e^\uparrow
\begin{pmatrix}
u\\
0\\
0\\
v\e{-\i\gamma}\\
\end{pmatrix}
\e{\i q\cos\theta_s x} 
+ t_e^\downarrow
\begin{pmatrix}
0\\
u\\
-v\e{-\i\gamma}\\
0\\
\end{pmatrix}
\e{\i q\cos\theta_s x}
+ t_h^\uparrow
\begin{pmatrix}
0\\
-v\e{\i\gamma}\\
u\\
0\\
\end{pmatrix}
\e{-\i q\cos\theta_s x}
+ t_h^\downarrow
\begin{pmatrix}
v \e{\i\gamma}\\
0\\
0\\
u\\
\end{pmatrix}
\e{-\i q\cos\theta_s x}
\Bigg].
\end{align}
\end{widetext}
The elements entering in the wave-functions above describing the quasiparticles read 
\begin{align}
a &= \frac{1}{\sqrt{1+[M_{xy}/(M+M_z)]^2}},\; b &= \frac{aM_{xy}}{M+M_z},
\end{align}
for the F part, while the superconducting coherence factors read
\begin{align}
u &= \sqrt{\frac{1}{2} + \frac{\sqrt{E^2 - \Delta^2}}{2E}},\notag\\
v &= \sqrt{\frac{1}{2} - \frac{\sqrt{E^2 - \Delta^2}}{2E}}.
\end{align}
We denote the F phase by $\phi$ and S phase by $\gamma$. Note that $\text{tan}\phi = -M_y/M_x$, such that 
the physical interpretation of the F phase is directly related to the direction of the magnetization in 
the $xy$-plane characterized by the azimuthal angle. An incoming electron with spin-$\uparrow$ is 
described by $\{s_\uparrow=1,s_\downarrow=0\}$ while a spin-$\downarrow$ electron is given by $\{s_\uparrow=0,s_\downarrow=1\}$. For convenience, we also introduce $S = s_\uparrow\cos\theta + s_\downarrow\cos\theta_A$, $\tilde{S} = s_\uparrow\cos\theta_A + s_\downarrow\cos\theta$. The 
boundary conditions for these wave-functions read 
\begin{align}\label{eq:boundary}
&\psi_\text{F}(0,y) = \Psi_\text{S}(0,y),\notag\\
&\frac{\Psi'_\text{S}(x,y)|_{x=0}}{2m_\text{S}} - 
\frac{\psi'_\text{F}(x,y)|_{x=0}}{2m_\text{F}} = V_0 - V_s\eta, 
\end{align}
where $\eta = (1,-1,1,-1)^\text{T}$ and $'$ denotes derivation with respect to $x$. Translational 
invariance along the $\hat{\mathbf{y}}$-direction implies conservation of the momentum $k_y$. This 
allows us to determine $\theta_s$ and $\theta_A$ as follows
\begin{align}
(s_\uparrow k^\uparrow + s_\downarrow k^\downarrow)\sin\theta &= q\sin\theta_s,\notag\\
(s_\uparrow k^\uparrow + s_\downarrow k^\downarrow)\sin\theta &= (s_\uparrow k^\downarrow + s_\downarrow k^\uparrow)\sin\theta_A.
\end{align}

\section{Presence of retroreflection}
\begin{figure}[h!]
\centering
\resizebox{0.47\textwidth}{!}{
\includegraphics{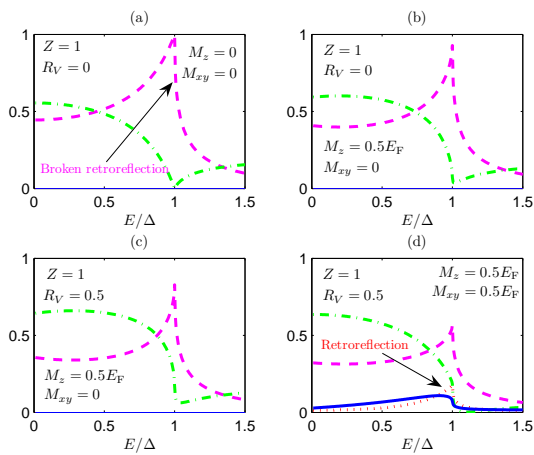}}
\caption{(Color online). Plot of the probability coefficients associated with the scattering processes at 
the interface. For an electron with incoming spin $\sigma$, the green (dash-dotted) line corresponds to 
normal reflection with spin $\sigma$, the magenta (dashed) line corresponds to Andreev-reflection of a hole 
with spin $-\sigma$, the blue (full) line designates reflection without branch-crossing with spin $-\sigma$,
while the presence of retroreflection, \ie Andreev-reflection of a hole with spin $\sigma$, is indicated
by the red (dotted) line. Note from (d) that in order to get retroreflection, both
an in-plane magnetization \textit{and} a spin-active barrier is required.}
\label{fig:proof}
\end{figure}

Several cases may now be studied, such as different effective masses in the F and S 
part, Fermi-vector mismatch, and the presence of a spin-active barrier. Solving 
Eq. (\ref{eq:boundary}) for the wave-functions in Eqs. (\ref{eq:wave}), one is able to 
obtain explicit expressions for the reflection coefficients of the scattering problem. 
This amounts to solving for $16$ unknown coefficients, and their derivation may be found 
in Appendix A. While the expressions for their amplitudes are quite cumbersome, their 
phase-dependences are simple and illustrate the new physics. In Tab. \ref{tab:phase}, we 
provide this phase-dependence for the cases of incoming $\uparrow$ and $\downarrow$ 
electrons \footnote{In general, there is also a contribution 
$\e{-\i[\text{arccos($E/\Delta$)}]}$ for $E<\Delta$, but this is 
irrelevant for the present discussion.}. It is seen that a coupling between $\phi$ and 
$\gamma$ is present in the phase of the hole with the same spin $\sigma$ as the incident 
electron. Ordinarily, retro-reflection is absent in the Andreev-scattering process at the
F/S junction such that the reflected hole and the incident electron carry opposite spins. 
However, it is clear from Tab. \ref{tab:phase} that were a hole with spin $\sigma$ to 
be generated in the scattering process, it would carry information about both the F 
and S phases. We interpret this as {\it induced spin-triplet pairing correlations} in 
the S part of the system, along with an electron-hole correlation in the ferromagnet.  
\begin{table}[h!]
\centering{
\caption{Phase-dependence of reflection coefficients. Here, 
"1" means that the quantity is real. An interplay between 
$\gamma$ and $\phi$ occurs when retro-reflection is present.}
	\label{tab:phase}
	\vspace{0.15in}
	\begin{tabular}{ccccc}
	  	 \hline
	  	 \hline
		 \textbf{Refl. coeff.}	\hspace{0.1in}	& $r_h^\uparrow$ \hspace{0.1in}& $r_h^\downarrow$ \hspace{0.1in}& $r_e^\uparrow$ \hspace{0.1in}& $r_e^\downarrow$  \\
	  	 \hline
	  	 Inc. spin-$\uparrow$	\hspace{0.1in} & $\e{-\i(\phi+\gamma)}$ \hspace{0.1in} & $\e{-\i\gamma}$ \hspace{0.1in} & 1 	 \hspace{0.1in} & $\e{-\i\phi}$		\\
	  	 Inc. spin-$\downarrow$ \hspace{0.1in} \hspace{0.1in} & $\e{-\i\gamma}$ \hspace{0.1in} & $\e{\i(\phi-\gamma)}$ \hspace{0.1in} & $\e{\i\phi}$ 	\hspace{0.1in} & 1\\
	  	 \hline
	  	 \hline
	\end{tabular}}
\end{table}
\\
\indent Although the phase-dependence of the reflection coefficients displayed in Tab. \ref{tab:phase} is 
intriguing, it remains to be demonstrated that the \textit{amplitudes} of these coefficients are non-zero. 
To illustrate that this is so, consider Fig. \ref{fig:proof} where we have plotted the probability coefficients 
[that differ from the reflection coefficents by a pre-factor, see Eq. (\ref{eq:probcof})] for normal incidence 
$\theta=0$; their derivation may be found in Appendix A. In (a), we have no exchange energy and a purely 
non-magnetic interfacial resistance, from which the result of Ref.~\onlinecite{btk} is reproduced. In (b), 
we have allowed for an exchange energy $M_z = 0.5E_\text{F}$, which results in a reduction of the 
Andreev-reflection amplitude. This is a consequence of the reduced carrier density of the spin-$\downarrow$ 
band due to the presence of a magnetic exchange energy. In the extreme limit of a completely spin-polarized ferromagnet, $M_z = E_\text{F}$, the subgap conductance is completely absent since there are no charge 
carriers in the spin-$\downarrow$ band at Fermi level. In (c), we also incorporated the effect of a magnetic 
scattering potential in the interfacial resistance, which is seen to slightly reduce the probability of the Andreev-reflection at $E=\Delta$. The novel features of the F/S junction are now presented in (d). When we 
allow for both a magnetic scattering potential \textit{and} local spin-flip processes in the form of a 
planar component of the magnetization, it it seen that retroreflection is established. In other words, a 
new transport channel is opened up for both spin and charge, namely reflected hole-like excitations with 
the same spin as the incoming electron. Note that the inclusion of this process is absent in most of the 
literature treating F/S junctions so far \cite{jong, zutic00, kashiwaya1999}.

\begin{figure}[h!]
\centering
\resizebox{0.5\textwidth}{!}{
\includegraphics{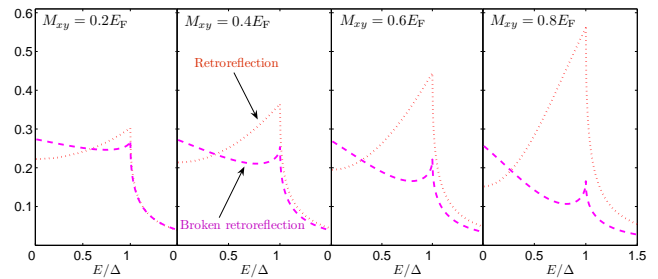}}
\caption{(Color online). Plot of probability coefficients for $Z=1$ and $R_V = 0.95$ in the absence of any net polarization for several values of $M_{xy}$. It is seen that for increasing $M_{xy}$, \ie larger effect of spin-flip scattering, the retroreflection process dominates the "normal" Andreev-reflection. }
\label{fig:study}
\end{figure}

To investigate how large the magnitude of the retroreflection coefficient may become, possibly even 
outgrowing the probability for "normal" Andreev-reflection, we plotted the case of zero net polarization 
for several values of $M_{xy}$ in Fig. \ref{fig:study}. It is seen that as $M_{xy}$ increases, the 
probability for retroreflection grows, and eventually becomes much larger than the probability for 
ordinary Andreev-reflection. Thus, for a tunneling junction with a barrier that discriminates 
significantly between spin-$\uparrow$ and spin-$\downarrow$ electrons, the presence of spin-flip 
processes may induce a substantial modification to the traditional picture of broken retroreflection.

Having established the presence of retroreflection, the next step is the consideration of how retroreflection 
leaves its signatures in experimentally measurable quantities. In this paper, we investigate how the presence 
of retroreflection may leave an experimental signature manifested in the conductance spectrum of a F/S 
junction. Although this shall be our focus, we note in passing that the reflection coefficients derived in 
Appendix A may also be used for the purpose of obtaining the current-voltage characteristics, spin-current, 
and spin conductance of the F/S junction. Normally, the charge- and spin-current may be written as
\begin{align}
j_\text{charge} = -e\sum_\sigma j^\sigma,\; j_\text{spin} = \sum_\sigma \sigma j^\sigma,
\end{align}
where $j^\sigma$ is the particle-current of electrons with spin $\sigma$ over the interface. However, in 
the presence of spin-flip scattering, defining a proper spin-current requires a more careful analysis 
\cite{shi}. One can always write down a well-defined spin-current in terms of physical spin-transport 
across the junction, but it may be very hard to experimentally distinguish whether the spin accumulation 
on either side of the interface should be attributed to physical spin transport or local spin-flip 
processes. The latter are present in \eg systems with significant spin-orbit coupling or an in-plane 
magnetic field with respect to the quantization axis, which results in scattering between the two 
spin bands. Accordingly, in this paper we will concern ourselves with the charge-current and the resulting 
conductance spectrum.

\section{Results}
In our theory, we have included the possibility of having a spin-active barrier, Fermi-vector mismatch, 
arbitrary strength of the exchange energy on the F side, and different effective masses in the two 
systems. Thus, we believe our model should be able to capture many essential and realistic features 
of a F/S junction that pertain to both interfacial properties as well as bulk effects on the F and 
S side, respectively. Since the case of easy-axis magnetization has been thoroughly investigated, 
we shall be mainly concerned with the presence of retroreflection, {\it which requires both 
spin-flip processes and a barrier acting as a spin-filter}. \\
\indent The single-particle tunneling conductance may be calculated by using the 
BTK-formalism \cite{btk}, and reads
\begin{align}\label{eq:G}
G(E) &= \sum_\sigma G^\sigma(E),\notag\\
G^\sigma(E) &= \int^{\pi/2}_{-\pi/2} \text{d}\theta \cos\theta P^\sigma G^\sigma(E,\theta),\notag\\
G^\sigma(E,\theta) &= G_N^{-1}[1 + R_h^\uparrow(E,\theta) + R_h^\downarrow(E,\theta)\notag\\
&\;\;- R_e^\uparrow(E,\theta) - R_e^\downarrow(E,\theta)],\notag\\
G_N &= \int^{\pi/2}_{-\pi/2} \text{d}\theta \cos\theta \frac{4\cos^2\theta}{4\cos^2\theta + Z^2},
\end{align}
where $Z = 2m_\text{F} V/k_F$ and $G_N$ is the tunneling conductance for a N/N junction. Note that r.h.s. of the equation for $G^\sigma(E,\theta)$ appears to be independent of $\sigma$. However, it is implicitly understood in this notation that the reflection coefficients appearing on the r.h.s. have been solved for an incoming electron with spin $\sigma$, and these differ in the cases $\sigma=\uparrow$ and $\sigma=\downarrow$ since the wavefunction is different [see Eq. (\ref{eq:wave})]. The different 
probabilities for having spin injection $\sigma$ in the presence of a net polarization is accounted 
for by the factor $P^\sigma = (1 + \sigma M_Z/E_F)/2$. The quantities $\{R_h^\sigma, R_e^\sigma\}$ 
are the probability coefficients for normal- and Andreev-reflection, and will be derived below. Note 
that these are \textit{not} in general equal to the square amplitude of the scattering coefficients, 
and in particular not so in this case. To see this, consider the current-density of probability 
$\mathbf{J}_\text{inc}$ that is incident on the barrier, 
\begin{align}
\mathbf{J}_\text{inc} = \frac{1}{2m_\text{F}\i}(\psi^*\nabla\psi - \psi\nabla\psi^*),
\end{align}
obeying the conservation law
\begin{equation}
\frac{\partial P}{\partial t} + \nabla\cdot \mathbf{J}_\text{inc} = 0.
\end{equation}
Here, $P = |\psi|^2$. Consulting Eq. (\ref{eq:wave}) and extracting the part of $\psi$ that 
corresponds to the incident wave-function, one readily obtains
\begin{equation}
\mathbf{J}_\text{inc} = \frac{\cos\theta}{m_\text{F}}(s_\uparrow k^\uparrow + s_\downarrow k^\downarrow) \hat{\mathbf{x}}.
\end{equation}
Since probability must be conserved, we have 
\begin{equation}\label{eq:conserve}
\mathbf{J}_\text{inc} = -\mathbf{J}_\text{refl} + \mathbf{J}_\text{trans},
\end{equation}
where the reflected probability current-density reads
\begin{align}
\mathbf{J}_\text{refl} &= \frac{1}{2m_\text{F}\i}[ (\psi_e^*\nabla\psi_e - \text{h.c.}) - (\psi_h^*\nabla\psi_h - \text{h.c.})],\notag\\
\psi_e &= r_e^\uparrow\begin{pmatrix}
a\\
b\e{-\i\phi}\\
\end{pmatrix}
\e{-\i k^\uparrow S x}
 + r_e^\downarrow\begin{pmatrix}
 -b\e{\i\phi}\\
 a\\
 \end{pmatrix}
\e{-\i k^\downarrow \tilde{S} x},\notag\\
\psi_h &= r_h^\uparrow\begin{pmatrix}
a\\
b\e{\i\phi}\\
\end{pmatrix}
\e{\i k^\uparrow S x}
 + r_h^\downarrow\begin{pmatrix}
 -b\e{-\i\phi}\\
 a\\
 \end{pmatrix}
\e{\i k^\downarrow \tilde{S} x}.
\end{align}
The opposite signs of the electron- and hole-part of $\psi$ entering $\mathbf{J}_\text{refl}$ pertain 
to the fact that their energy eigenvalues have opposite signs, as one may infer from the BdG-equations 
that are used to derive the explicit expression for $\mathbf{J}_\text{refl}$ from 
Eq. (\ref{eq:conserve}). One finds that
\begin{equation}
\mathbf{J}_\text{refl} = -\frac{1}{m_\text{F}}[ k^\uparrow S |r_e^\uparrow|^2 + k^\uparrow S |r_h^\uparrow|^2 + k^\downarrow \tilde{S} |r_e^\downarrow|^2 + k^\downarrow \tilde{S} |r_h^\downarrow|^2]\hat{\mathbf{x}}.
\end{equation}
The same procedure may now be applied to $\mathbf{J}_\text{trans}$, such that Eq. (\ref{eq:conserve}) 
can be written as 
\begin{equation}\label{eq:probconserve}
1 = \sum_\sigma (R_e^\sigma + R_h^\sigma + T_e^\sigma + T_h^\sigma)
\end{equation}
upon division with $|\mathbf{J}_\text{inc}|$. From this, one infers that 
\begin{align}\label{eq:probcof}
R_e^\uparrow &= |r_e^\uparrow|^2 \frac{k^\uparrow S}{s_\uparrow k^\uparrow \cos\theta + s_\downarrow k^\downarrow \cos\theta},\notag\\
R_e^\downarrow &= |r_e^\downarrow|^2 \frac{k^\downarrow \tilde{S}}{s_\uparrow k^\uparrow \cos\theta + s_\downarrow k^\downarrow \cos\theta},\notag\\
R_h^\uparrow &= |r_h^\uparrow|^2 \frac{k^\uparrow S}{s_\uparrow k^\uparrow \cos\theta + s_\downarrow k^\downarrow \cos\theta},\notag\\
R_h^\downarrow &= |r_h^\downarrow|^2 \frac{k^\downarrow \tilde{S}}{s_\uparrow k^\uparrow \cos\theta + s_\downarrow k^\downarrow \cos\theta}.
\end{align}
The coefficients $\{R_e^\sigma, R_h^\sigma, T_e^\sigma, T_h^\sigma\}$ have the status of probability 
coefficients for their respective processes, and obey the conservation law Eq. (\ref{eq:probconserve}). 
Note that in the absence of exchange splitting, \ie F$\to$ N and $\theta_A = \theta$, one obtains 
$R_i^\sigma = |r_i^\sigma|^2$.

\subsection{Effect of Fermi-vector mismatch}
\begin{figure}[h!]
\centering
\resizebox{0.35\textwidth}{!}{
\includegraphics{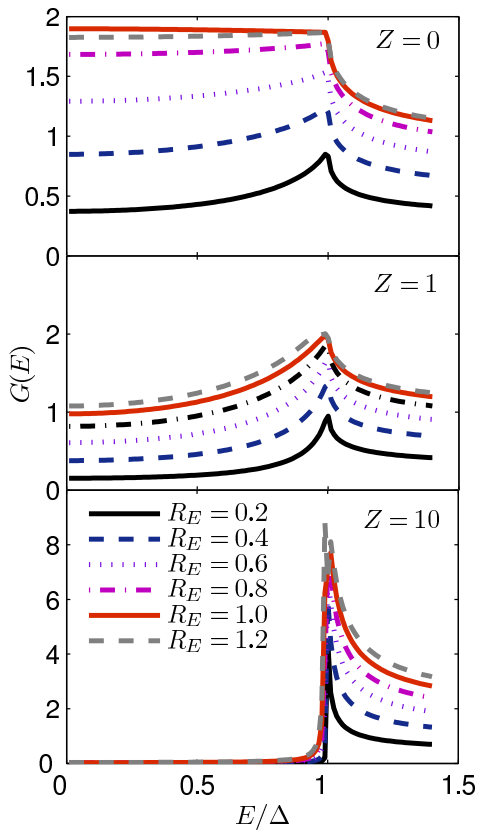}}
\caption{(Color online). Conductance spectrum for weak spin-flip scattering 
($M_{xy} = 0.1E_F$) and $M_z = 0$ with $R_V = 0.5$ for several values of $Z$.}
\label{fig:FWM1}
\end{figure}

To account for the Fermi-vector mismatch, we introduce a parameter $R_E = E_\text{S}/E_\text{F}$. This 
allows the Fermi energies in the F and S regions to be different, which effectively models unequal carrier 
densities and bandwidths on each side of the junction. For ferromagnet/high-$T_c$ superconductor junctions, 
an appropriate choice appears to be \cite{zutic00} $R_E \leq 1$. In our study, however, we will consider 
values of $R_E$ both less than and greater than unity. To begin with, we fix the strength of the planar 
contribution to the exchange energy at $M_{xy} = 0.1E_F$ and set $M_z = 0$, plotting the conductance 
spectrum for several values of $R_E$. We fix the ratio $R_V = V_s/V_0 = 0.5$, such that the conditions 
for retroreflection are fulfilled. For each figure, we consider zero ($Z=0$), weak ($Z=1$), and large 
($Z=10$) interfacial resistance; $Z=0$ corresponds to the point-contact (also called metallic contact, 
in some literature) while $Z\to\infty$ equivalents the tunneling limit. The conductance spectrum for 
weak spin-flip scattering ($M_{xy} = 0.1E_F$) and $M_z = 0$ with $R_V = 0.5$ for several values of 
$Z$, is depicted in Fig. \ref{fig:FWM1}. From Fig. \ref{fig:FWM1}, we infer that the conductance 
behaves in a monotonic way upon variation of $R_E$, and that the conductance is suppressed with 
decreasing $R_E$.

\begin{figure}[h!]
\centering
\resizebox{0.35\textwidth}{!}{
\includegraphics{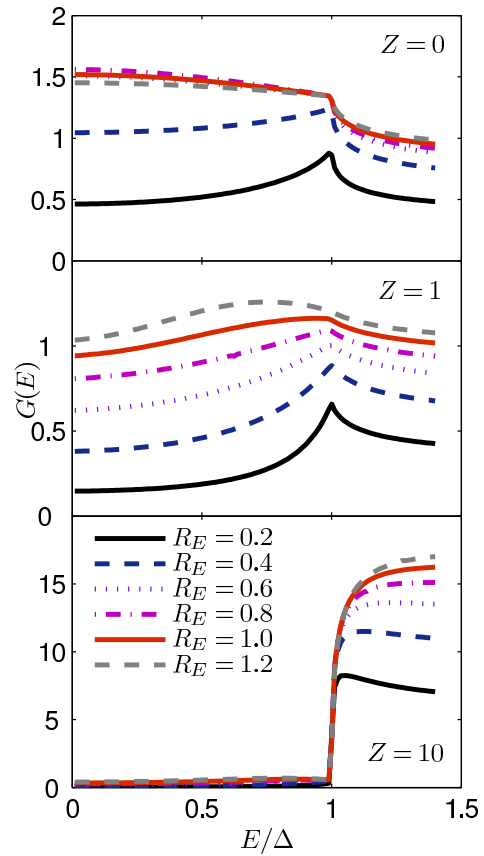}}
\caption{(Color online). Conductance spectrum for strong spin-flip scattering ($M_{xy} = 0.5E_F$) and $M_z = 0$ with a strongly spin-dependent barrier ($R_V = 0.95$) for several values of $Z$.}
\label{fig:FWM2}
\end{figure}

Next, we increase the exchange energy to $M_{xy} = 0.5E_F$ and set $R_V = 0.95$ such that spin-flip processes become more dominant and the barrier discriminates strongly between spin-$\uparrow$ and spin-$\downarrow$ electrons. The resulting $G(E)$ is illustrated in Fig. \ref{fig:FWM2}, where it is seen that a nonmonotonic behaviour appears. Specifically, the peak at $E=\Delta$ vanishes for $R_E \simeq 1$, as is most clearly seen for the case of large interfacial resistance. 

\begin{figure}[h!]
\centering
\resizebox{0.35\textwidth}{!}{
\includegraphics{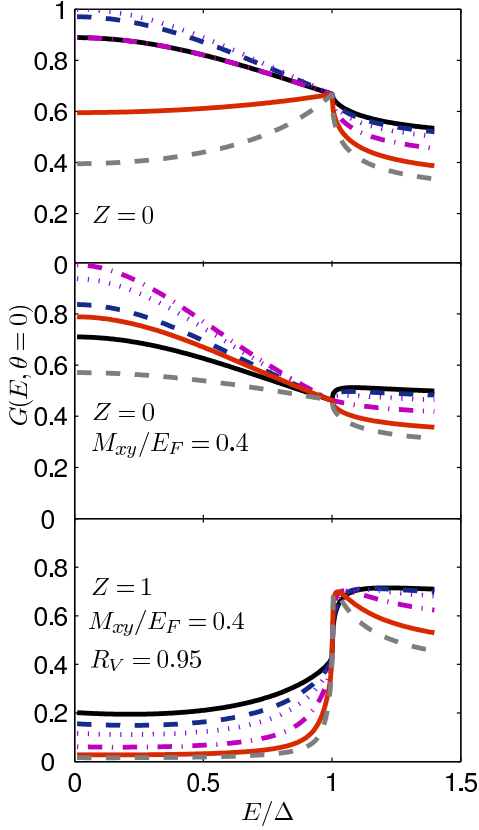}}
\caption{(Color online). Conductance spectrum for zero spin-flip scattering and purely 
non-magnetic scattering potential (upper panel), spin-flip scattering and purely non-magnetic 
scattering potential (middle panel), spin-flip scattering and mixed magnetic/non-magnetic 
scattering potential. For all panels, $M_z/E_F = 0.866$ for comparison with 
Ref.~\onlinecite{zutic00}. The lines are given at $E=1.4$ for the upper panel as 
follows (from top to bottom): $R_E = \{1, 1/\sqrt{2}, 1/2, 1/4, 1/9, 1/16\}$.}
\label{fig:effectFWM}
\end{figure}

One of the results of Refs.~\onlinecite{zutic99, zutic00} was that the effect of Fermi-vector mismatch 
yielded an increased subgap conductance when there was a net spin-polarization. As an important consequence, 
this finding suggested that the interfacial barrier parameter $Z$ was not sufficient to account for the 
conductance features in the presence of both spin polarization and Fermi-vector mismatch, since the 
increase of subgap conductance could not be reproduced by varying $Z$ alone. In Figs. \ref{fig:FWM1} 
and \ref{fig:FWM2}, no such increase in subgap conductance was found, but these correspond to an 
unpolarized case since $M_z = 0$. In order to investigate how the spin-flip scattering and spin-active 
barrier affects this particular feature of the Fermi-vector mismatch, we plot the normal incidence 
$\theta=0$ conductance $G(E,\theta=0)$ for the same parameters as Fig. 1 in 
Refs.~\onlinecite{zutic99, zutic00} for the sake of direct comparison. Note that due to a different 
scaling of the conductance to make it dimensionless, the quantitative results for $G(E,\theta=0)$ is 
not the same as the result in Refs.~\onlinecite{zutic99, zutic00}, although the qualitative aspect is 
identical. This is because we scale the conductance on $G_N$ given by Eq. (\ref{eq:G}) For $Z=0$, this 
merely amounts to a factor of 2. In the upper panel of Fig. \ref{fig:effectFWM}, we reproduce Fig. 1b 
of Ref.~\onlinecite{zutic00} to illustrate our consistency with their results. Note that the parameter 
$L_0^2$ in Ref.~\onlinecite{zutic00} is equivalent to our $R_E$ when $R_m = 1$, \ie the effective 
masses are the same. The middle panel now includes spin-flip scattering with $M_{xy} = 0.4E_F$, while 
$Z=0$. The lower panel shows the combined effect of planar magnetization and a spin-active barrier, 
resulting in triplet correlations, with $M_{xy} = 0.4E_F$ and $\{Z=1, R_V = 0.95\}$. It is seen that 
the qualitative change is most dramatic when the conditions for retroreflection are fulfilled.

\subsection{Effect of exchange energy}
\begin{figure}[h!]
\centering
\resizebox{0.35\textwidth}{!}{
\includegraphics{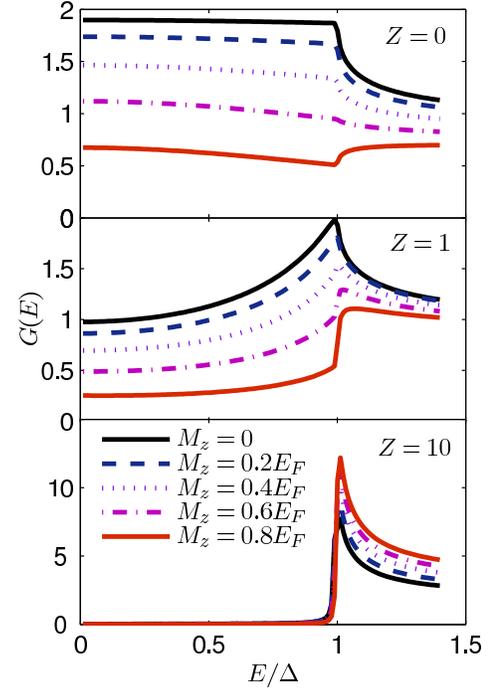}}
\caption{(Color online). Conductance spectra for various non-magnetic scattering potentials 
upon varying the polarization of the ferromagnet with $M_{xy} = 0.1E_F$ and $R_V = 0.5$.}
\label{fig:EX1}
\end{figure}

We now proceed to consider how the strength of the exchange energy, both planar ($M_{xy}$) and 
easy-axis ($M_z$), affects the conductance spectrum. We  set the masses and Fermi energies 
to be equal in the F and S part of the system, and study how the angularly averaged $G(E)$ is 
affected by increasing $M_Z$ for a given $M_{xy}$. Let us first set $M_{xy} = 0.1E_F$ and 
$R_V = 0.5$, as shown in Fig. \ref{fig:EX1}. In accordance with our previous observation 
that Andreev-reflection is inhibited by a net polarization in the F part of the system, it 
is seen that the conductance is suppressed with increasing $M_z$. However, in the lower panel 
of Fig. \ref{fig:EX1} where the tunneling limit of the junction is considered, the conductance 
increases with $M_z$ for $E>\Delta$. 

\begin{figure}[h!]
\centering
\resizebox{0.35\textwidth}{!}{
\includegraphics{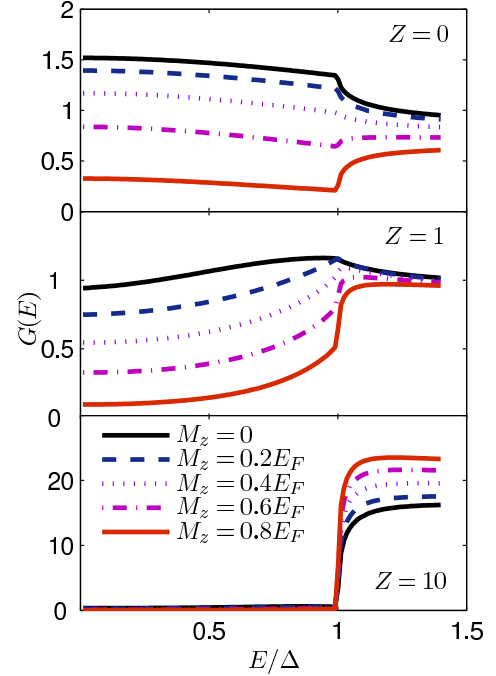}}
\caption{(Color online). Conductance spectra for various non-magnetic scattering potentials upon 
varying the polarization of the ferromagnet with $M_{xy} = 0.5E_F$ and $R_V = 0.95$.}
\label{fig:EX2}
\end{figure}

Increasing the strength of the spin-flip scattering and also the spin-dependence of the barrier, 
the resulting conductance spectra are shown in Fig. \ref{fig:EX2} with $M_{xy} = 0.5E_F$ and 
$R_V = 0.95$. The general effect of optimizing the conditions for the presence of retroreflection 
processes seems to be a "smoothing out" of the conductance: the sharp features at $E=\Delta$ 
become blunt, an observation which is most clearly revealed in the tunneling limit. As an 
experimental consequence, the nature of the features at $E=\Delta$ in the case of a high-resistance 
interface could thus offer information concerning to what degree retroreflection is present in 
the system.

\subsection{Effect of different effective masses}
\begin{figure}[h!]
\centering
\resizebox{0.35\textwidth}{!}{
\includegraphics{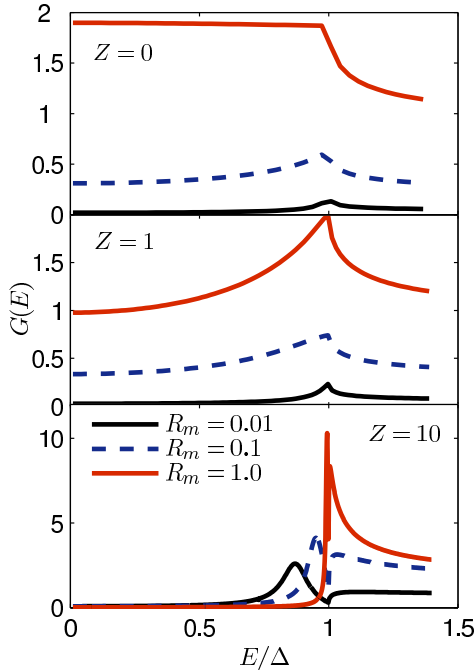}}
\caption{(Color online). Conductance spectra for different effective masses with parameters 
$M_{xy} = 0.1E_F$ and $R_V = 0.5$.}
\label{fig:M1}
\end{figure}

\begin{figure}[h!]
\centering
\resizebox{0.35\textwidth}{!}{
\includegraphics{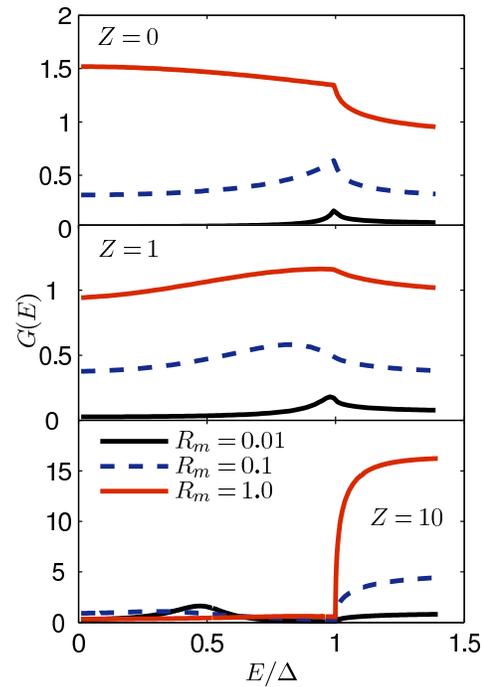}}
\caption{(Color online). Conductance spectra for different effective masses with parameters $M_{xy} = 0.5E_F$ and $R_V = 0.95$.}
\label{fig:M2}
\end{figure}

To investigate the effect of different effective masses in the F and S part of the system, we consider 
three ratios: $R_m = \ms/\mf \in \{0.01,0.1,1\}$. In Fig. \ref{fig:M1}, we have plotted the case of 
weak spin-flip scattering and a moderate spin-dependence of the barrier, while in Fig. \ref{fig:M2} 
we investigate significant spin-flip scattering and a strongly spin-dependent interfacial resistance. 
In the first case, decreasing $R_m$ clearly inhibits the tunneling conductance with no exotic features 
present except the usual peak at $E=\Delta$. In the tunneling limit, it is interesting to observe that 
only in the case $R_m = 1$ is the maximum of the conductance located at $E=\Delta$. Upon decreasing 
$R_m$, one sees that the characteristic peak of the spectrum is translated to lower energies and that 
it becomes less sharp. There is still a sudden increase of current at $E=\Delta$, manifested as a 
jump in the conductance spectrum, but it is less protruding for lower ratios of $R_m$ than unity. 

When the conditions for retroreflection become more pronounced, as is the case in Fig. \ref{fig:M2}, 
one may again observe the general modification of the conductance to a more featureless curve in the 
case of no barrier and a weak barrier ($Z=1$), as was the case in the previous subsection. In the 
tunneling limit, the presence of retroreflection also modifies the spectra such that the sharp peak 
is lost at the gap energy, although the sudden jump due to the initiated flow of current at 
$E=\Delta$ is still there.

\subsection{Effect of magnetic and non-magnetic scattering potential}
\begin{figure}[h!]
\centering
\resizebox{0.35\textwidth}{!}{
\includegraphics{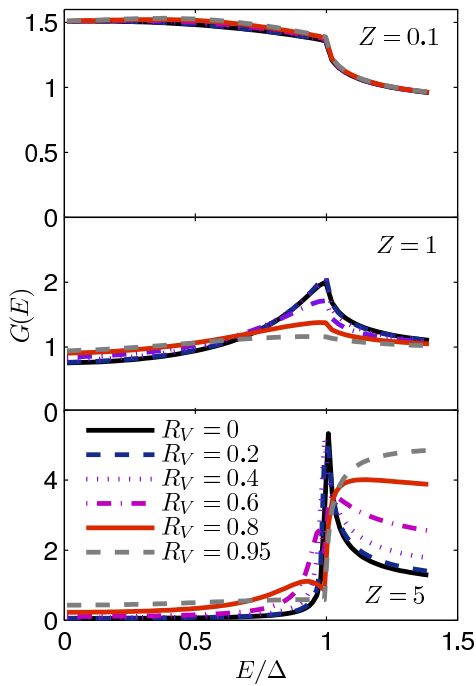}}
\caption{(Color online). Conductance spectra in the presence of retroreflection but in the 
absence of any net polarization. Here, $M_{xy} = 0.5E_F$ while $M_Z =0$.}
\label{fig:B1}
\end{figure}

In this section, we show that the conductance spectrum may reveal clear-cut signatures of the 
presence of retro-reflection as a result of the interplay between $V_0$ and $V_s$ when 
$M_{xy}\neq0$. We keep the latter fixed at $M_{xy}=0.5E_F$, and plot $G(E)$ for $Z\in \{0.1,1,5\}$ 
while varying the strength of the magnetic scattering potential. From Fig. \ref{fig:B1}, we see 
that at $Z=0.1$, the presence of retro-reflection is very weak and the conductance spectrum 
remains virtually unaltered as $V_s$ is varied. At $Z=1$, the effect of increasing the strength 
of the magnetic potential of the barrier, acting as a spin-filter, corresponds to a reduction of 
the conductance-peak at $E=\Delta$. This is in agreement with our previous observations that the 
presence of retroreflection appears to have a smoothing effect on the conductance spectrum, causing 
it to soften its characteristic features. At $Z=5$, the crossover from a sharp peak at $E=\Delta$ 
at small $R_V$ to a "waterfall"-shape for large $R_V$ is clearly illustrated. We suggest that 
this signature could be used as a feature that unveils the presence of retroreflected holes in 
the system, and thus indicates triplet correlations due to the interplay between spin-flip 
processes and a barrier acting as a spin-filter.

\begin{figure}[h!]
\centering
\resizebox{0.35\textwidth}{!}{
\includegraphics{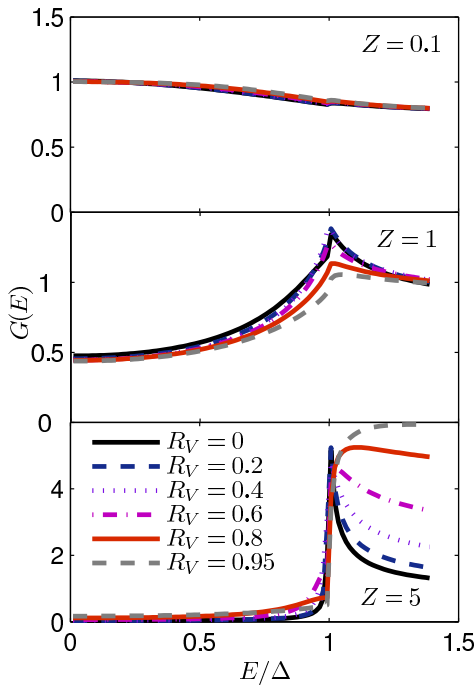}}
\caption{(Color online). Conductance spectra in the presence of retroreflection and a net 
polarization. Here, $M_{xy} = 0.5E_F$ while $M_Z =0.5E_F$.}
\label{fig:B2}
\end{figure}

To investigate how a net polarization will affect the conductance spectra in this case, consider 
Fig. \ref{fig:B2} which illustrates the conductance for the same parameters as in Fig. \ref{fig:B1} 
except that now $M_z = 0.5E_F$. In agreement with previous remarks, the conductance suffers a 
general reduction due to the net polarization in the upper and lower panel. However, the conductance 
corresponding to the strongest polarization comes out on top in the tunneling limit as in the case 
of Fermi-vector mismatch. Apart from this, the same features as in Fig. \ref{fig:B1} are present, 
with retroreflection leaving its fingerprint most obviously in the behaviour of the conductance 
at $E=\Delta$ in the tunneling limit. 

\section{Discussion}
\begin{figure}[h!]
\centering
\resizebox{0.35\textwidth}{!}{
\includegraphics{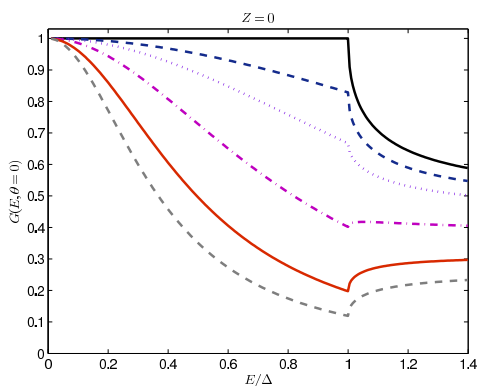}}
\caption{(Color online). In the limit $M_{xy} \to 0$, the formation of a ZBCP is observed with 
decreasing $R_E$. This illustrates how the effect of Fermi-vector mismatch may "mimick" the 
usual signature of unconventional superconductivity, namely the appearance of a ZBCP for 
certain crystal orientations. This was first discussed in  Ref.~\onlinecite{zutic00}, see 
their Fig. 3. From top to bottom, the curves correspond to the following pairs of $(R_E, M_z/E_F)$: (1,0), $(\frac{1}{\sqrt{2}}, \frac{1}{\sqrt{2}})$, $(\frac{1}{2}, 0.866)$, $(\frac{1}{4}, 0.968)$, $(\frac{1}{9}, 0.994)$, and $(\frac{1}{16}, 0.998)$.}
\label{fig:comparison}
\end{figure}

We have shown that the presence of a spin-active barrier combined with a planar component of the 
magnetization in the F induces new features in the proximity effect in a F/S junction. Physically, 
this may be understood by realizing that only an $S_z=0$ triplet component is induced for a 
spin-active barrier in the absence of spin-flip processes near the junction, while the equal-spin 
$(S_z=\pm1)$ triplet components are generated only if a spin-flip potential is also present. On 
the other hand, spin-flip processes alone in the absence of a spin-active barrier would inhibit 
singlet pairing without generating any triplet components. An interesting opportunity that arises 
due to the restoration of retro-reflection is the fact that one may generate currents with a varying 
degree of spin-polarization in the F part. In the conventional case, an incident electron with spin 
$\sigma$ is reflected as either an electron with spin $\sigma$ or hole with spin $-\sigma$ in these 
systems. In the present case, however, the reflected electrons and holes may carry \textit{either} 
$\uparrow$ and $\downarrow$ spin, depending on parameters such as the magnitude of the exchange 
energy and the intrinsic/spin-dependent barrier strength. In principle, it could possible to generate 
pure spin-currents without charge-currents and vice versa as a result of the additional allowed spin 
state of the reflected holes and electrons. It is also intriguing to observe that due to the coupling 
between $\phi$ and $\gamma$, it may be possible to obtain a Josephson current in a S/F/S hybrid structure
that is sensitive to a rotation of the magnetization in the ferromagnetic part, which has been recently discussed in Refs.~\onlinecite{kreuzy, pajovic}. \\
\indent It was shown in Ref.~\onlinecite{bergeretPRL} that if a local inhomogeneity of the 
magnetization in the vicinity of a F/S interface was present, a spin-triplet component of the S order 
parameter will be generated and penetrate into the F much deeper than the spin-singlet component. In a 
S/half-metal/S junction, it has been found that S triplet correlations would be induced on both 
sides of the junctions in the presence of spin-mixing and spin-flip scattering at the interfaces 
\cite{eschrigPRL} (see also Ref.~\onlinecite{tokuyasu}). We have found that spin-triplet pairing 
correlations may be induced in the presence of a spin-active barrier, \ie intrinsic spin-mixing 
at the interface, and a planar magnetization relative to the quantization axis. It seems reasonable 
to suggest that these findings are closely related to the conditions put forward by 
Ref.~\onlinecite{eschrigPRL}, since planar magnetization components may effectively act as a 
spin-flip scattering potential. Our results are thus consistent with the findings of recent 
studies, although we have adressed several new aspects of the scattering problem in the present 
paper. In particular, we have found an interplay between in-plane magnetization direction
and superconducting phase which to our knowledge has not been investigated before. 
Moreover, we compute detailed conductance spectra of the F/S junction under many different 
conditions. \\
\indent One of the important findings of Refs.~\onlinecite{zutic99, zutic00} was that a 
zero-bias conductance peak (ZBCP) would develop under the right conditions in the F/S 
junction, and the effect was attributed to the influence of Fermi-vector mismatch. Usually, 
the appearance of a ZBCP is associated with unconventional superconductivity where it may 
appear due to the different phases felt by the transmitted electron-like and hole-like 
quasiparticles in the superconductor \cite{tanaka}. However, Zutic and Valls 
\cite{zutic99, zutic00} showed that no unconventional superconductivity was required 
to obtain a ZBCP, and that the effect of Fermi-vector mismatch in a F/S junction thus 
offered a different mechanism for the formation of a ZBCP than the usual one, attributed 
to a $\vk$-dependent gap. However, it should be noted that the ZBCP obtained in 
Refs.~\onlinecite{zutic99, zutic00} is not as sharp (delta-function like) as the ZBCP depicted 
in \eg Ref.~\onlinecite{tanaka}, where unconventional superconductors (high $T_c$ 
$d$-wave, to be specific) were considered. \\
\indent In the present paper, we consider a more general situation than Zutic and Valls, 
allowing for a completely arbitrary magnetization direction and a spin-active barrier. 
As we have shown, this changes the physical picture dramatically and opens up a new 
transport channel for both charge and spin, namely retroreflected holes. For consistency, 
we show that we are able to completely reproduce Fig. 3 of Ref.~\onlinecite{zutic00}, where 
the conductance for normal incidence $\theta=0$ is presented 
(our Fig. \ref{fig:comparison}). 

In contrast to Zutic and Valls\cite{zutic00}, due to the unwieldy expressions for the reflection 
coefficients (see Appendix A), we are not able to give analytically the condition that yields the 
largest value of the conductance at zero-bias [cf. their Eq. (3.4)]. It is thus not straight-forward 
to identify  the proper parameter regime that would yield the maximum value of $G(0)$. We therefore 
leave the question concerning how spin-flip scattering and a spin-active barrier affect 
the formation of a ZBCP in a F/S-junction, as open. \\
\indent Scattering on the barrier leads to a suppression of the S order parameter close [of order 
coherence length, ${\cal{O}}(\xi)$] to the junction. For a weakly polarized ferromagnet, we expect 
that inclusion of a spatial variation of the order parameter does not change our results 
qualitatively, since it is well-known that the approximation of a constant order parameter up to 
the junction is excellent in a N/$s$-wave superconductor junction (see \eg Ref.~\onlinecite{bruder}). 
For a strongly polarized ferromagnet, the superconducting singlet order parameter may however be 
suppressed significantly in the vicinity of the gap \cite{eschrigPRL}. For unconventional pairing 
symmetries ($d$-wave), it was shown in Ref.~\onlinecite{tanaka2000} that the effect of taking into 
account the suppression of the order parameter in the presence of Andreev bound surface states remains 
almost unchanged around zero bias voltage, although a broadening of the ZBCP is observed. Since no 
zero energy surface states are present for a pure $s$-wave singlet component of the superconducting 
order parameter, we believe that our approximation of a step-function $\Delta$ should be justified. \\
\indent It is worth noting that a F/S junction as considered here with a spin-active barrier is in 
some respects similar to previously studied F/F/S junctions \cite{kikuchi} if the magnetization 
directions of the two F layers are non-collinear. While Ref.~\onlinecite{kikuchi} considers the conductance 
spectrum in the case of collinear magnetization directions of the F layers, a previous study \cite{kopu} 
has developed a quite general framework for dealing with F/S junctions by introducing a phenomenological 
spin-mixing angle which describes a spin-active interface. In Ref.~\onlinecite{kopu}, the conductance is 
explicitly calculated for a half-metallic ferromagnet/$s$-wave superconductor junction. In the present paper, 
we have developed a similar framework for treating F/S junctions with a spin-active interface, but using 
a different formalism. Our theory allows for describing a very wide range of physical phenomena, such as 
arbitrary magnetization strength/direction of the ferromagnet, a spin-active barrier, Fermi-vector mismatch, 
and different effective masses in the two systems. 
We have explicitly computed the conductance spectra for the metallic case with non-collinear magnetizations 
between the F-part and the spin-active barrier in a F/S system. Hence, our work expands on the results of
Ref. \onlinecite{kikuchi} and Ref. \onlinecite{kopu}, and we reproduce their results in the
appropriate limits. \\
\indent The similarity of our model with F/F/S junction with noncollinear magnetizations may be understood 
by realizing that using a spin basis that diagonalizes 
the scattering matrix of one ferromagnet will cause the magnetization in the other ferromagnet to 
effectively look like a spin flip term and vice versa. Although this analogy could be of some use 
for comparing the present system under consideration with F/F/S junctions, it should not be taken 
too far since in our case we are dealing with an insulating, very thin barrier with both magnetic 
and non-magnetic scattering potentials as opposed to a conducting ferromagnetic layer. \\
\indent Another issue that deserves mentioning is that the magnetic field due to the magnetization 
of the F will penetrate into the thin-film structure of the S along the plane. An in-plane magnetic 
field may actually coexist uniformly \cite{meservey} with $s$-wave S in a thin-film (in contrast to 
the bulk case \cite{varma, shen}) and effects such as orbital pair-breaking or formation of vortices 
will be prohibited as long as the thickness $t$ of the film is less than both $\lambda$ and $\xi_0$. 
It is also reasonable to neglect any exchange interactions in the S since the induced field due to 
the magnetization is much smaller [of order ${\cal{O}}$($10^{-3}$)] than the exchange field in the 
F, and can thus be safely neglected \cite{buzdinRMP}. Moreover, we stress that the clean limit has 
been considered in the present paper, which hopefully provides an initial idea of the physics that 
can be expected when the effect of disorder is included in the system, although this requires a 
separate analysis.\\

\section{Summary}
In this paper, we have presented a detailed investigation of the conductance spectra of a F/S junction, 
expanding previous work substantially by allowing for a completely arbitrary direction of magnetization, 
which effectively accounts for spin-flip scattering due to a planar component of the magnetization, and 
a spin-active barrier. Our procedures amounts to an extension of the BTK-formalism, along the lines of 
several other workers (\eg \cite{tanaka, kashiwaya1999}), and have given us the advantadge of obtaining 
analytical solutions, primarily due to the step-function approximation for the superconducting and 
magnetic order parameters. \\
\indent From our results, one may infer that several new qualitative features arise due to the presence 
of spin-flip scattering and a spin-active barrier. We demonstrate the 
re-entrance of retroreflection for the Andreev-reflected hole, which is absent for an easy-axis ferromagnet 
with a purely non-magnetic interfacial scattering potential. This opens up a new transport channel for both 
spin and charge, and is interpreted as a signature of spin-triplet correlations in the system. In this 
context, a most interesting interplay between the superconducting phase $\gamma$ and the planar magnetization orientation characterized by the azimuthal angle $\phi$ arises in the phase coherence of retroreflected 
holes. This particular feature may be exploited in terms of a Josephson current in a S/F/S junction that 
responds to a rotation of $\phi$. \\
\indent As our main result, we have investigated the influence on the conductance spectra due to different effective 
masses, Fermi-vector mismatch, strength of the exchange energy, and the influence of varying the relative 
strength of magnetic and non-magnetic scattering in the F/S junction. Our findings are consistent with 
those of Ref.~\onlinecite{zutic00} with respect to the observation of an increased subgap conductance for 
increasing Fermi-vector mismatch for a large spin polarization. In the presence of a spin-active barrier, 
however, this effect vanishes. The general influence of retroreflection on the conductance spectra seems 
to be a softening of the sharp features such as peaks and dips at $E=\Delta$. Also, as a signature which 
should be clearly discernable experimentally, a crossover from peak to "waterfall" shape takes place 
in the tunneling limit at the gap energy.\\
\indent We believe that our angle of approach for treating the F/S junction in the extended BTK-formalism 
should suffice to shed light on the rich physics and concomitant important phenomena that are present in 
such systems, which is of particular relevance in the context of spin polarized tunneling spectroscopy.

\section*{Acknowledgments}
\noindent J. L. acknowledges useful discussions with M. Gabureac. This work was supported by the Norwegian Research Council Grant Nos. 158518/431, 158547/431, (NANOMAT), 
and 167498/V30 (STORFORSK).

\appendix

\section{Derivation of scattering coefficients}\label{sec:app}

From the boundary conditions, the condition of continuity of the wave-function yields the
expressions
\begin{align}\label{eq:r}
&s_\uparrow a - s_\downarrow b \e{-\i\phi} + r_e^\uparrow - r_e^\downarrow b\e{\i\phi} = t_e^\uparrow u + t_h^\downarrow v\e{\i\gamma},\notag\\
&s_\uparrow b\e{-\i\phi} + s_\downarrow a + r_e^\uparrow b\e{-\i\phi} + r_e^\downarrow a = t_e^\downarrow u - t_h^\uparrow v\e{\i\gamma}\notag\\
&r_h^\uparrow a - r_h^\downarrow b\e{-\i\phi} = -t_e^\downarrow v\e{-\i\gamma} + t_h^\uparrow u,\notag\\
&r_h^\uparrow b\e{\i\phi} + r_h^\downarrow a = t_e^\uparrow v\e{-\i\gamma} + t_h^\downarrow u,
\end{align}
while the matching of derivatives at $x=0$ yields
\begin{align}
&(V_0-V_s)(t_e^\uparrow u + t_h^\downarrow v\e{\i\gamma}) = \frac{\i q\cos\theta_s}{2m_\text{S}}(ut_e^\uparrow - v\e{\i\gamma}t_h^\downarrow) \notag\\
&- \frac{\i}{2m_\text{F}}[\cos\theta(k^\uparrow s_\uparrow a -  k^\downarrow s_\downarrow b \e{\i\phi}) - k^\uparrow S a r_e^\uparrow + k^\downarrow \tilde{S} b \e{\i\phi} r_e^\downarrow],\notag\\
&(V_0+V_s)(t_e^\downarrow u -t_h^\uparrow v\e{\i\gamma}) = \frac{\i q\cos\theta_s}{2m_\text{S}}(ut_e^\downarrow + v\e{\i\gamma}t_h^\uparrow)\notag\\
&-\frac{\i}{2m_\text{F}}[\cos\theta(k^\uparrow s_\uparrow b\e{-\i\phi} + k^\downarrow s_\downarrow a) - k^\uparrow Sb\e{-\i\phi}r_e^\uparrow - k^\downarrow\tilde{S} a r_e^\downarrow],\notag\\
&(V_0-V_s)(-t_e^\downarrow v\e{-\i\gamma} + t_h^\uparrow u) = -\frac{\i q\cos\theta_s}{2m_\text{S}}(v\e{-\i\gamma} t_e^\downarrow + u t_h^\uparrow) \notag\\
&- \frac{\i}{2m_\text{F}}(k^\uparrow S a r_h^\uparrow - k^\downarrow \tilde{S} b\e{-\i\phi} r_h^\downarrow),\notag\\
&(V_0+V_s)(t_e^\uparrow v\e{-\i\gamma} + t_h^\downarrow u) = \frac{\i q\cos\theta_s}{2m_\text{S}}(t_e^\uparrow v\e{-\i\gamma} - t_h^\downarrow u) \notag\\
&-\frac{\i}{2m_\text{F}}(k^\uparrow S b\e{\i\phi} r_h^\uparrow + k^\downarrow\tilde{S} a r_h^\downarrow).
\end{align}
Solving for the transmission coefficients, one is left with the reduced set of equations
\begin{align}\label{eq:4syst}
t_e^\uparrow A_1 + t_e^\downarrow B_1 \e{\i\phi} + t_h^\uparrow C_1 \e{\i(\phi+\gamma)} + t_h^\downarrow D_1 \e{\i\gamma} &= X_1,\notag\\
t_e^\uparrow A_2\e{-\i\phi} + t_e^\downarrow B_2 + t_h^\uparrow C_2 \e{\i\gamma} + t_h^\downarrow D_2 \e{\i(\gamma-\phi)} &= X_2,\notag\\
t_e^\uparrow A_3\e{-\i(\phi+\gamma)} + t_e^\downarrow B_3 \e{-\i\gamma} + t_h^\uparrow C_3  + t_h^\downarrow D_3 \e{-\i\phi} &= 0,\notag\\
t_e^\uparrow A_4 \e{-\i\gamma} + t_e^\downarrow B_4 \e{\i(\phi-\gamma)} + t_h^\uparrow C_4 \e{\i\phi} + t_h^\downarrow D_4 &= 0.
\end{align}
From Eqs. (\ref{eq:4syst}), one finds that
\begin{align}\label{eq:t}
t_h^\downarrow &= X_1 F_1 \e{-\i\gamma} + X_2 F_2 \e{\i(\phi-\gamma)},\notag\\
t_h^\uparrow &= X_2 R_1 \e{-\i\gamma} + R_e t_h^\downarrow \e{-\i\phi},\notag\\
t_e^\downarrow &= P_1 t_h^\uparrow\e{\i\gamma} + P_2 t_h^\downarrow\e{\i(\gamma-\phi)},\notag\\
t_e^\uparrow &= -(B_4t_e^\downarrow\e{\i\phi} + C_4t_h^\uparrow \e{\i(\phi+\gamma)} + D_4 t_h^\downarrow \e{\i\gamma})/A_4,
\end{align}
such that the reflection coefficients $\{r_h^\sigma, r_e^\sigma\}$ may be obtained by back-substitution of Eqs. (\ref{eq:t}) into Eqs. (\ref{eq:r}). We have defined the following auxiliary quantities:
\begin{align}
X_1 &= \frac{1}{2m_\text{F}}(k^\uparrow \cos\theta s_\uparrow a - k^\downarrow\cos\theta s_\downarrow b\e{\i\phi} \notag\\
&+ k^\uparrow S a s_\uparrow - k^\downarrow \tilde{S} s_\downarrow \e{\i\phi}),\\
X_2 &= \frac{1}{2m_\text{F}}(k^\uparrow \cos\theta b\e{-\i\phi} + k^\downarrow \cos\theta s_\downarrow a \notag\\
&+ k^\uparrow S s_\uparrow \e{-\i\phi} + k^\downarrow \tilde{S} s_\downarrow a),\\
F_1 &= [D_1 + C_1R_2 + P_1B_1R_2 + B_1P_2 \notag\\
&- \frac{A_1}{A_4}(B_4P_2 + B_4P_1R_2 + R_2C_4 + D_4)]^{-1},\\
F_2 &= F_1[\frac{A_1}{A_4}(B_4P_1R_1 + R_1C_4) - B_1P_1R_1 - C_1R_1],\\
R_1 &= [C_2 + B_2P_1 - \frac{A_2}{A_4}(B_4P_1 + C_4)]^{-1},\\
P_1 &= (\frac{C_4A_3}{A_4} - C_3)/(B_3 - \frac{A_3B_4}{A_4}),\\
R_2 &= R_1[B_2P_2 + D_2 -\frac{A_2}{A_4}(B_4P_2 + D_4)],\\
P_2 &= (\frac{D_4A_3}{A_4} - D_3)/(B_3 - \frac{A_3B_4}{A_4}),
\end{align}
in addition to
\begin{widetext}
\begin{align}
A_1 &= \i(V_0-V_s)u + \frac{1}{2\ms}q\cos\theta_s u + \frac{u}{2\mf}(k^\uparrow S a^2 + k^\downarrow \tilde{S} b^2),\; A_2 = \frac{1}{2\mf}(k^\uparrow S - k^\downarrow\tilde{S})abu,\\
A_3 &= \frac{1}{2\mf}(k^\downarrow \tilde{S} - k^\uparrow S)abv,\; A_4 = \i(V_0+V_s)v + \frac{1}{2\ms}q\cos\theta_sv - \frac{v}{2\mf}(k^\uparrow S b^2 + k^\downarrow \tilde{S} a^2),\\
B_1 &= \frac{1}{2\mf}(k^\uparrow S - k^\downarrow\tilde{S})abu,\; B_2 = \i(V_0+V_s)u + \frac{1}{2\ms}q\cos\theta_su + \frac{u}{2\mf}(k^\uparrow Sa^2 + k^\downarrow\tilde{S}b^2),\\
B_3 &= -\i(V_0 - V_s)v - \frac{1}{2\ms}q\cos\theta_s v+ \frac{v}{2\mf}(k^\uparrow S a^2 + k^\downarrow\tilde{S} b^2),\; B_4 = -\frac{1}{2\mf}(k^\downarrow\tilde{S} - k^\uparrow S)abv,\\
C_1 &= \frac{1}{2\mf}(k^\downarrow\tilde{S} - k^\uparrow S)abv,\; C_2 = -\i(V_0+V_s)v + \frac{1}{2\ms}q\cos\theta_s v - \frac{v}{2\mf}(k^\uparrow Sb^2 + k^\downarrow\tilde{S}a^2),\\
C_3 &= \i(V_0-V_s)u - \frac{1}{2\ms}q\cos\theta_s u - \frac{u}{2\mf}(k^\uparrow S a^2 + k^\downarrow\tilde{S}b^2),\; C_4 = -\frac{1}{2\mf}(k^\uparrow S - k^\downarrow\tilde{S})abu,\\
D_1 &= \i(V_0-V_s)v - \frac{1}{2\ms}q\cos\theta_s v + \frac{v}{2\mf}(k^\uparrow S a^2 + k^\downarrow\tilde{S}b^2),\; D_2 = \frac{1}{2\mf}(k^\uparrow S - k^\downarrow\tilde{S})abu,\\
D_3 &= \frac{1}{2\mf}(k^\downarrow\tilde{S} - k^\uparrow S)abu,\; D_4 = \i(V_0+V_s)u - \frac{1}{2\ms}q\cos\theta_s u - \frac{u}{2\mf}(k^\uparrow Sb^2 + k^\downarrow\tilde{S} a^2)
\end{align}
\end{widetext}

\end{document}